\documentclass[a4paper,11pt]{article}
\usepackage{amsmath,amsthm,graphicx}
\usepackage[english]{babel}
\usepackage[width=0.9\textwidth,indention=5mm,justification=centerlast,
 font=small,labelfont=bf,labelsep=period]{caption}
\usepackage{hyperref}

\advance\textwidth 24mm  \advance\oddsidemargin -12mm
\advance\textheight 20mm \advance\topmargin -15mm

\theoremstyle{plain}
\newtheorem{statement}{Statement}

\def\a{\alpha}
\def\b{\beta}
\def\g{\gamma}
\def\la{\lambda}
\def\const{\mathop{\rm const}}
\def\SP<#1>{\langle#1\rangle}

\begin{document}
\title{On vector analogs of the modified Volterra lattice}
\author{V.E. Adler\thanks{L.D. Landau Institute for Theoretical Physics,
 Chernogolovka, Russia. E-mail: adler@itp.ac.ru},\quad
 V.V. Postnikov\thanks{Sochi Branch of Peoples' Friendship University of Russia,
 Sochi, Russia.\newline \strut~~~~~E-mail: postnikovvv@rambler.ru}}

\date{1 August 2008}

\maketitle

\begin{abstract}
Modified Volterra lattice admits two vector generalizations. One of them is
studied for the first time. The zero curvature representations, B\"acklund
transformations, nonlinear superposition principle and the simplest explicit
solutions of soliton and breather type are presented for both vector lattices.
The relations with some other integrable equations are established.
\medskip

\end{abstract}

\section{Introduction}

Vector equations are an important and rather well studied class of
integrable system. Among others, we mention but a few works in this area
\cite{Fordy, Tsuchida_Wadati, Sokolov_Wolf} containing the examples and
classification results for the vectorial systems of derivative nonlinear
Schr\"odinger type which are in some relation to the theme of our paper.
There are also several interesting results for the vector
differential-difference equations, or lattices, see e.g.
\cite{Ablowitz_Ohta_Trubatch, Tsuchida}, but this field seems less
investigated. The aim of our work is the study of the vector lattices
\begin{gather}
\label{V1}
 V_{n,x}=2\SP<V_n,V_{n+1}-V_{n-1}>V_n-\SP<V_n,V_n>(V_{n+1}-V_{n-1}),\\[0.5em]
\label{V2}
 V_{n,x}=\SP<V_n,V_n>(V_{n+1}-V_{n-1}),
\end{gather}
which define two integrable generalizations of the very well known modified
Volterra lattice. Equation (\ref{V1}) was introduced in
\cite{Adler_Svinolupov_Yamilov} among the other examples of the
multi-component lattices related to Jordan algebraic structures. Second
lattice is considered here for the first time, up to our knowledge, despite
of its more simple form.

The main tool in the study of a nonlinear integrable equation is its
representation as the compatibility condition for auxiliary linear systems.
In the differential-difference setting this method was developed in the
classical papers \cite{Case_Kac, Manakov}. In our paper we restrict
ourselves by the version of dressing method based on Darboux-B\"acklund
transformations and their nonlinear superposition principle. The main
results for the scalar lattice are given in Section \ref{s.sc}. The main
body of the paper, Sections \ref{s.V1}, \ref{s.V2}, contains generalizations
of this method for both vector lattices (\ref{V1}), (\ref{V2}), as well as
the simplest explicit solutions of soliton and breather type.

A characteristic feature of integrability is the consistency of the equation
with an infinite hierarchy of other equations. In particular, B\"acklund
transformations define the discrete part of this hierarchy and lead to the
discrete equations on the square grid. Usually, one starts this way from the
continuous equations of KdV type, however an understanding appeared recently
that the lattice equations of Volterra type lead to the same result as well
\cite{Nijhoff_Hone_Joshi, Adler_Suris, Levi_Petrera_Scimiterna}. This
relation has not been observed in the vector case yet, although the discrete
equation related to the lattice (\ref{V1}) has been introduced in the paper
\cite{Schief}, see also \cite{Bobenko_Suris_2002, Bobenko_Suris_2005,
Adler_1995}.

The continuous part of the picture (Section \ref{s.sym}) is more
traditional. It was observed in works of Levi \cite{Levi} and Shabat,
Yamilov \cite{Shabat_Yamilov} that integrable Volterra type lattices define
a special kind of B\"acklund transformations for equations of nonlinear
Schr\"odinger type. This remains valid for the vector analogs as well. The
connection with a two-dimensional lattice relative to the Volterra lattice
introduced by Mikhailov \cite{Mikhailov} is of interest, too. Finally, it
should be noted that the approach based on the continuous symmetries is the
most effective one in the classification problem of integrable equations,
both continuous and discrete \cite{Mikhailov_Shabat_Yamilov}. The complete
classification of scalar Volterra type lattices was obtained by Yamilov
\cite{Yamilov_1983} by use of the symmetry approach, see also
\cite{Levi_Yamilov, Yamilov_2006}. Some progress in classification of vector
equations and lattices has been achieved recently \cite{Sokolov_Wolf,
Meshkov_Sokolov, Tsuchida_Wolf, Adler_2008}. We discuss some open problems
in this field in the concluding Section \ref{s.more}.

\section{Scalar case}\label{s.sc}

\subsection{Zero curvature representations}\label{s.sc.ZCR}

The following notations for the auxiliary linear equations are used throughout
the paper:
\begin{equation}\label{sc.LAM}
 \Psi_{n+1}=L_n\Psi_n,\quad
 \Psi_{n,x}=A_n\Psi_n,\quad
 \tilde\Psi_n=M_n\Psi_n.
\end{equation}
Modified Volterra lattice
\begin{equation}\label{sc.vx}
 v_{n,x}=(v^2_n+a)(v_{n+1}-v_{n-1})
\end{equation}
is equivalent to the compatibility condition $L_{n,x}=A_{n+1}L_n-L_nA_n$
with the matrices
\begin{equation}\label{sc.LA}
 L_n=\begin{pmatrix}
  \dfrac{a}{\la} & v_n \\[1em]
  -v_n & \la \end{pmatrix},\quad
 A_n=\begin{pmatrix}
  \dfrac{a^2}{\la^2}+v_{n-1}v_n  & \dfrac{a}{\la}v_n+\la v_{n-1}\\[1em]
  -\dfrac{a}{\la}v_{n-1}-\la v_n & \la^2+v_{n-1}v_n \end{pmatrix}.
\end{equation}
The Darboux-B\"acklund transformation is defined by the matrix
\begin{equation}\label{sc.M}
 M_n=\frac1{a+\mu^2f^2_n}\begin{pmatrix}
  \mu(a^2-\mu^2\la^2)-a\mu(\la^2-\mu^2)f^2_n & -(a^2-\mu^4)\la f_n \\[0.5em]
  (a^2-\mu^4)\la f_n & a\mu(\la^2-\mu^2)-\mu(a^2-\mu^2\la^2)f^2_n \end{pmatrix}.
\end{equation}
Moreover, the compatibility condition $\tilde L_nM_n=M_{n+1}L_n$ is
equivalent to the pair of discrete Riccati equation for the variable $f_n$:
\begin{equation}
\label{sc.vvff}
 v_n=\frac{\mu f_{n+1}-af_n/\mu}{1+f_nf_{n+1}},\qquad
 \tilde v_n=\frac{\mu f_n-af_{n+1}/\mu}{1+f_nf_{n+1}}
\end{equation}
and the condition $M_{n,x}=\tilde A_nM_n-M_nA_n$ completes this system with
the continuous Riccati equation
\begin{equation}
\label{sc.vvf}
 f_{n,x}=\Bigl(\frac{a}{\mu}v_{n-1}+\mu v_n\Bigr)f^2_n
  +\Bigl(\frac{a^2}{\mu^2}-\mu^2\Bigr)f_n+\mu v_{n-1}+\frac{a}{\mu}v_n.
\end{equation}
Notice also that the variable $f_n$ satisfies, in virtue of equations
(\ref{sc.vvff}), (\ref{sc.vvf}), the lattice
\begin{equation}\label{sc.fx}
 f_{n,x}=\frac{(\mu^2+af^2_n)(a+\mu^2f^2_n)(f_{n+1}-f_{n-1})}
  {\mu^2(1+f_{n+1}f_n)(1+f_nf_{n-1})}.
\end{equation}
Starting from a known solution $v_n$ of the lattice (\ref{sc.vx}) the common
solution of the first equation (\ref{sc.vvff}) and equation (\ref{sc.vvf})
is constructed by the formula $f_n=\phi_n/\varphi_n$ where
$\Psi=(\phi,\varphi)$ is a particular solution of two first equations
(\ref{sc.LAM}) at $\la=\mu$. Then the second equation (\ref{sc.vvff})
defines the new solution $\tilde v_n$.

\begin{figure}[t]
\centerline{
\includegraphics[width=7cm]{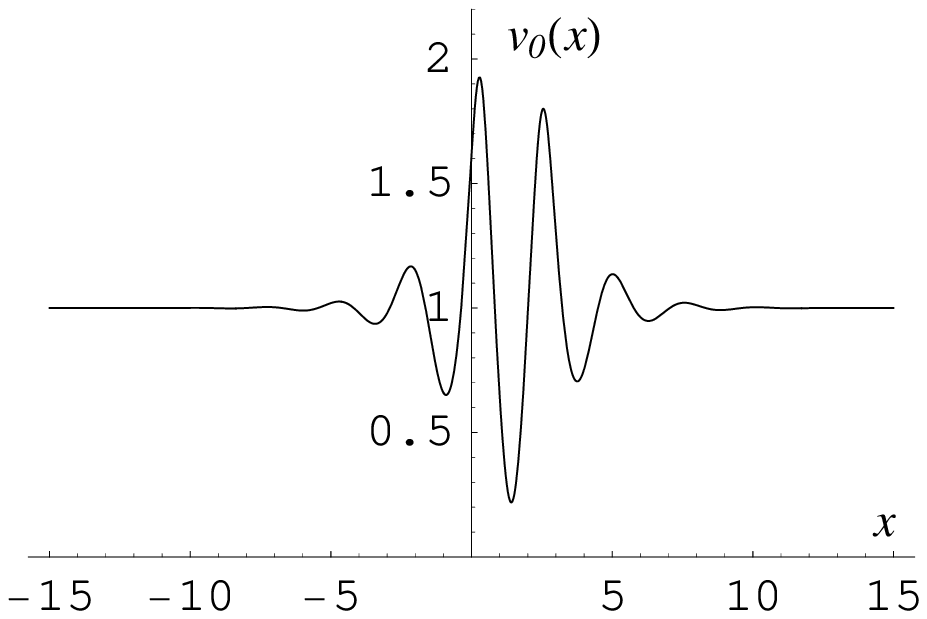}\qquad
\includegraphics[width=7cm]{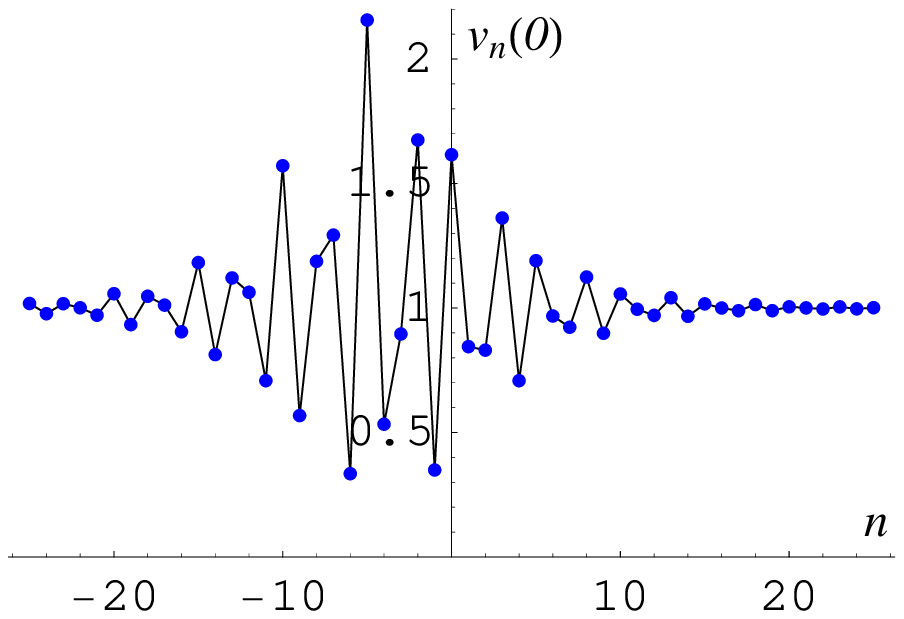}}
\centerline{\includegraphics[width=10cm]{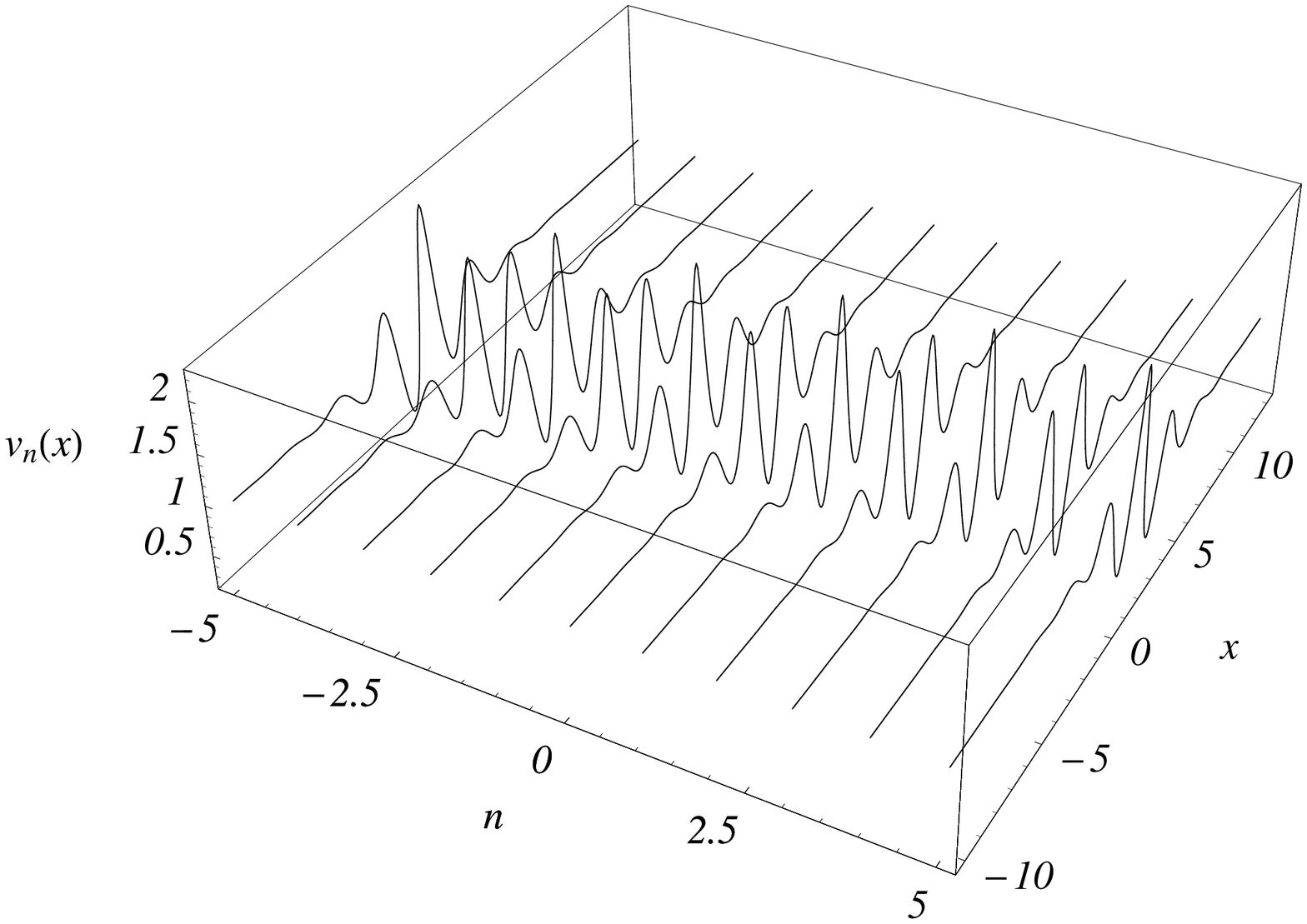}}
\caption{Breather of the lattice (\ref{sc.vx}); $a=1$,
 $\g^{(1)}_1=\bar\g^{(2)}_1=0.5+1.5i$, $k^{(1)}=k^{(2)}=1$.}
\label{fig:sc.breather}
\end{figure}

For example, in order to construct solutions of soliton type one takes
$v_n=1$ as the seed solution (obviously, the choice of another constant is
equivalent to scaling of parameter $a$; some generalization can be achieved
via dressing of blinking solution $v_{2n}=\a$, $v_{2n+1}=\b$). The
eigenvalues of the matrix $L_n|_{v_n=1,\la=\mu}$ are defined from the
equations
\[
 \g_1+\g_2=\mu+a/\mu,\quad \g_1\g_2=1+a
\]
and the corresponding solution of the linear equations is
\[
 \varphi_n=\g^n_1e^{(\g^2_1+2)x}+k\g^n_2e^{(\g^2_2+2)x},\quad
 \phi_n=\mu\varphi_n-\varphi_{n+1}
\]
(we do not consider the case of multiple roots $\g_1=\g_2$ which leads to
rational in $n,x$ solutions). The ratio $f_n=\phi_n/\varphi_n$ defines the
solution of the lattice (\ref{sc.fx}) of kink type (provided $\g_1/\g_2>0$,
$k>0$) and the substitution into the second equation (\ref{sc.vvff}) gives
the soliton of the lattice (\ref{sc.vx}). The construction of $N$-soliton
solution uses the set of particular solutions
$(\phi^{(j)}_n,\varphi^{(j)}_n)$ corresponding to the values of parameters
$\mu^{(j)}$, $k^{(j)}$, $j=1,\dots,N$. If $a>0$ then the lattice
(\ref{sc.vx}) admits the breather solutions corresponding to the pairs of
complex conjugated points in the discrete spectrum
($\mu^{(1)}=\bar\mu^{(2)}$, $k^{(1)}=\bar k^{(2)}$).

\subsection{Nonlinear superposition principle}\label{s.sc.NSP}

The direct recomputing of the variables $f$ is a more convenient way to
iterate the Darboux transformation than applying the matrices $M$ and
recomputing the wave functions. This leads to the nonlinear superposition
principle of Darboux transformations in the form of some Yang-Baxter mapping
\cite{Adler_Bobenko_Suris_2004}. Let the variables $f^{(j)}_n$ be
constructed from the particular solutions of the linear systems at
$\mu=\mu^{(j)}$, and let $f^{(j,j_1,\dots,j_s)}_n$ denote the variables
obtained from $f^{(j)}_n$ by consequent application of Darboux
transforms with parameters $\mu^{(j_1)},\dots,\mu^{(j_s)}$. Then the
permutability of Darboux transformations is equivalent to the following
equality for the matrices of the form (\ref{sc.M}):
\[
  M(f^{(j,k,\sigma)}_n,\mu^{(j)})M(f^{(k,\sigma)}_n,\mu^{(k)})
 =M(f^{(k,j,\sigma)}_n,\mu^{(k)})M(f^{(j,\sigma)}_n,\mu^{(j)})
\]
where $\sigma$ stands for a tail sequence of distinct indices. This equation
is uniquely solvable with respect to $f^{(j,k,\sigma)}_n,f^{(k,j,\sigma)}_n$
and thus the mapping is defined
\begin{gather}
\nonumber
 \binom{f^{(j,\sigma)}_n}{f^{(k,\sigma)}_n} \mapsto
 \binom{f^{(j,k,\sigma)}_n}{f^{(k,j,\sigma)}_n}=
 \binom{R(f^{(j,\sigma)}_n,f^{(k,\sigma)}_n;\mu^{(j)},\mu^{(k)})}
       {R(f^{(k,\sigma)}_n,f^{(j,\sigma)}_n;\mu^{(k)},\mu^{(j)})},\\[0.5em]
\label{sc.R}
 R(f,g;\mu,\nu)=
 \frac{\mu\nu^3(\nu g-\mu f)-a\nu(\mu^2-\nu^2)fg^2-a^2(\mu g-\nu f)}
      {\mu\nu^3(\mu g-\nu f)g+a\nu(\mu^2-\nu^2)-a^2(\nu g-\mu f)g}.
\end{gather}

Another formulation of nonlinear superposition principle brings to a
discrete 4-point equation on the square grid for some new variable
$z^{(j,k)}_n$ (the subscript corresponds to the shift in the Volterra
lattice and is dummy, superscripts enumerate the Darboux transformations).
This equation is not too convenient for the purpose of the vector
generalizations which we have in mind, however it is of interest by itself
and we spend some space to describe it. The form of the equation depends on
the sign of $a$.

In the simplest case $a=0$ equations (\ref{sc.vvff}) imply the relation
\[
 \frac{\mu}{\tilde v_n}-\frac{\mu}{v_{n-1}}=f_{n+1}-f_{n-1}
\]
which allows to introduce the variable $z_n$ accordingly to the
equations
\[
 f_n=\mu(\tilde z_n-z_{n-1}),\qquad 1/v_n=z_{n+1}-z_{n-1}.
\]
This change turns the relations (\ref{sc.vvff}) into a single equation
\[
 (\tilde z_{n+1}-z_n)(z_{n+1}-\tilde z_n)=\mu^{-2}
\]
which define Darboux transformation in terms of the variable $z$. Now,
consider another Darboux transformation corresponding to the value
$\la=\nu$:
\[
 (\hat z_{n+1}-z_n)(z_{n+1}-\hat z_n)=\nu^{-2}.
\]
The easy calculation proves that the double Darboux transformations
coincide: $\hat{\tilde z}_n=\tilde{\hat z}_n$ and moreover, the common value
is given by the superposition formula:
\[
 (\hat{\tilde z}_n-z_n)(\hat z_n-\tilde z_n)=\mu^{-2}-\nu^{-2}.
\]
In other words, the Darboux transformations and superposition formula form
the triple which is 3D-consistent, or consistent around a cube
\cite{Adler_Bobenko_Suris_2003}. The iterations of Darboux transformation
bring to the discrete equation on the square grid (with fixed subscript)
\begin{equation}\label{sc.0NSP}
 (z^{(j+1,k+1)}_n-z^{(j,k)}_n)(z^{(j,k+1)}_n-z^{(j+1,k)}_n)
  =(\mu^{(j)})^{-2}-(\nu^{(k)})^{-2}.
\end{equation}
This is a very well-known 3D-consistent equation which defines as well the
nonlinear superposition principle of the classical Darboux transformation
for Schr\"odinger operator. This coincidence is not too surprising since it
is known for long that Volterra type lattices are symmetries of the dressing
chains which define B\"acklund transformations for KdV type equations (this
relation was discussed, from the different points of view, e.g. in
\cite{Shabat_Yamilov, Nijhoff_Hone_Joshi, Adler_Suris,
Levi_Petrera_Scimiterna}).

Analogously, in the case $a=-c^2$ the variable $z_n$ is introduced
accordingly to the formulae
\[
 f_n=\frac{\mu(\tilde z_n+z_{n-1})}{c(\tilde z_n-z_{n-1})},\qquad
 v_n=c\frac{z_{n+1}+z_{n-1}}{z_{n+1}-z_{n-1}}.
\]
After this the relations (\ref{sc.vvff}) turn into equation
\[
 a(\tilde z_{n+1}-z_n)(z_{n+1}-\tilde z_n)
 =-\mu^2(\tilde z_{n+1}+z_n)(z_{n+1}+\tilde z_n),
\]
and (\ref{sc.0NSP}) is replaced by equation
\begin{equation}\label{sc.-NSP}
 r(\mu^{(j)})\bigl(z^{(j,k)}_nz^{(j,k+1)}_n+z^{(j+1,k)}_nz^{(j+1,k+1)}_n\bigr)
 =r(\nu^{(k)})\bigl(z^{(j,k)}_nz^{(j+1,k)}_n+z^{(j,k+1)}_nz^{(j+1,k+1)}_n\bigr)
\end{equation}
where $r(\la)=(\la^2-a)/(\la^2+a)$, which is equivalent to nonlinear
superposition principle for $\sinh$-Gordon equation.

Finally, if $a=c^2$ then the change
\[
 f_n=\frac{\mu(1+z_{n-1}\tilde z_n)}{c(z_{n-1}-\tilde z_n)},\qquad
 v_n=c\frac{1+z_{n+1}z_{n-1}}{z_{n+1}-z_{n-1}}
\]
is used which brings equations (\ref{sc.vvff}) to the form
\[
 a(\tilde z_{n+1}-z_n)(z_{n+1}-\tilde z_n)
 =\mu^2(1+\tilde z_{n+1}z_n)(1+z_{n+1}\tilde z_n)
\]
and leads to equation
\begin{equation}\label{sc.+NSP}
\begin{gathered}
 \bigl(r(\mu^{(j)})+r(\nu^{(j)})\bigr)
 \bigl(z^{(j,k+1)}_n-z^{(j+1,k)}_n\bigr)
 \bigl(z^{(j,k)}_n-z^{(j+1,k+1)}_n\bigr)
\qquad\qquad\qquad \\ \qquad\qquad\qquad
 =\bigl(r(\mu^{(j)})-r(\nu^{(j)})\bigr)
 \bigl(1+z^{(j,k+1)}_nz^{(j+1,k)}_n\bigr)
 \bigl(1+z^{(j,k)}_nz^{(j+1,k+1)}_n\bigr)
\end{gathered}
\end{equation}
equivalent to nonlinear superposition principle for $\sin$-Gordon equation.
Equation (\ref{sc.-NSP}) turns into (\ref{sc.+NSP}) under the complex change
$z\to(i-z)/(i+z)$, so that these equations are two different real forms of
one and the same equation.

Concluding this Section, we notice that an analogous construction scheme
exists also for solutions of the Volterra lattice
\[
 u_{n,x}=u_n(u_{n+1}-u_{n-1}).
\]
The corresponding formulae are even much simpler, for example the equations
\[
 u_n=(v_n-\mu)(v_{n+1}+\mu),\quad \tilde u_n=(v_{n+1}-\mu)(v_n+\mu)
\]
replaces (\ref{sc.vvff}) while the role of the lattice (\ref{sc.fx}) is
played by the lattice (\ref{sc.vx}) at $a=-\mu^2$. Therefore, the lattice
(\ref{sc.fx}) is actually the second modification of Volterra lattice. This
sequence is analogous to the sequence of equations KdV $\to$ mKdV $\to$
$\exp$-CD which can be obtained by continuous limit from the lattices under
consideration. Unfortunately, although the Volterra lattice admits some
multi-component generalizations \cite{Salle,Suris}, the vector ones are
absent, this is why we have started from the more complicated object.

\section{First vector generalization}\label{s.V1}

Sometimes the zero curvature representation for a vector generalization can
be obtained just by passing to the block matrices. Unfortunately, this is
not the case for the matrices (\ref{sc.LA}), (\ref{sc.M}). It turns out,
however, that such block generalization is easy if one consider the linear
equations for the length 3 vector with the components consisting of the
products of the components of $\Psi$. Additionally, it is convenient to
apply a gauge transformation in order to make the determinants of the
matrices $L,M$ constant and the matrix $A$ traceless. In this way we come to
the following matrices which define, as can be easily verified, the zero
curvature representation for the lattice (\ref{sc.vx}) at $a=0$ and its
B\"acklund transformation:
\[
 L_n=\begin{pmatrix}
  0 & 0 & 1\\[0.7em]
  0 & -1 & \dfrac{\la}{v_n}\\[1em]
  1 & -\dfrac{2\la}{v_n} & \dfrac{\la^2}{v^2_n}
  \end{pmatrix},\qquad
 M_n=\begin{pmatrix}
  \dfrac{\la^2}{\mu^2f^2_n} & -\dfrac{2\la}{\mu f_n} & 1\\[1em]
  \dfrac{\la}{\mu f_n} & -1-\dfrac{\la^2}{\mu^2} & \dfrac{\la f_n}{\mu} \\[1em]
  1 & -\dfrac{2\la f_n}{\mu}  & \dfrac{\la^2f^2_n}{\mu^2}\end{pmatrix},
\]\[
 A_n=\la\begin{pmatrix}
  -\la & 2v_{n-1} & 0\\[0.5em]
  -v_n & 0 & v_{n-1} \\[0.5em]
  0 & -2v_n  & \la\end{pmatrix}.
\]
It is not a problem to find the matrices for the general case $a\ne0$, but
they are more cumbersome. Fortunately, we will not need them, since one of
the vector lattices exists only in the case $a=0$ anyway, and for the second
one this assumption does not lead to the loss of generality (see Section
\ref{s.V2}).

The block matrices for the vector lattices are derived from here under the
``proper'' interpretation of $v_n$ as a vector-valued quantity. To make
notation more clear we toggle to the upper case for the vectors. We assume
that the vector space is equipped with a symmetric scalar product
$\SP<U,V>=\SP<V,U>$. The identity operator is denoted $I$ and the linear
form $V^\top$, inverse vector $V^{-1}$ and operator $UV^\top$ are defined as
follows:
\[
 V^\top(U)=\SP<V,U>,\qquad V^{-1}=\frac{V}{\SP<V,V>}\qquad UV^\top(W)=U\SP<V,W>.
\]
In the case of finite-dimensional Euclidean space one can think of $V$ as of
the column vector and of $V^\top$ as of the row vector.

The first vector analog of the lattice (\ref{sc.vx}) exists only at $a=0$.
It is of the form \cite{Adler_Svinolupov_Yamilov}
\begin{equation}\label{V1.Vx}
 V_{n,x}=2\SP<V_n,V_{n+1}-V_{n-1}>V_n-\SP<V_n,V_n>(V_{n+1}-V_{n-1}).
\end{equation}
This lattice appears as the compatibility condition for the linear
systems
\begin{align}
\label{V1.T}
 T\begin{pmatrix} \psi_{n-1} \\ \Psi_n \\ \psi_n \end{pmatrix}
 &=\begin{pmatrix}
  0 & 0 & 1 \\
  0 & -I & \la V^{-1}_n \\
  1 & -2\la(V^{-1}_n)^\top & \la^2/\SP<V_n,V_n>
 \end{pmatrix}
 \begin{pmatrix} \psi_{n-1} \\ \Psi_n \\ \psi_n \end{pmatrix},\\[2mm]
\label{V1.D}
 D_x\begin{pmatrix} \psi_{n-1} \\ \Psi_n \\ \psi_n \end{pmatrix}
 &=\la\begin{pmatrix}
  -\la & 2V^\top_{n-1} & 0 \\
   -V_n & 0 & V_{n-1} \\
   0 & -2V^\top_n & \la
 \end{pmatrix}
  \begin{pmatrix} \psi_{n-1} \\ \Psi_n \\ \psi_n \end{pmatrix}.
\end{align}
Notice that the systems (\ref{V1.T}), (\ref{V1.D}) possess the first
integral in common
\begin{equation}\label{J}
 J=\SP<\Psi_n,\Psi_n>-\psi_n\psi_{n-1},\qquad (T-1)(J)=D_x(J)=0.
\end{equation}
The Darboux transformation is defined by a particular solution at the zero
level of this first integral.

\begin{statement}
Let $F_n=\Phi_n/\phi_n$ where $\psi=\phi$, $\Psi=\Phi$ is a particular
solution of the linear systems (\ref{V1.T}), (\ref{V1.D}) at $\la=\mu$ and
at $J=0$. Then the transformation
\begin{equation}\label{V1.DT}
 \begin{pmatrix}\tilde\psi_{n-1}\\[1mm] \tilde\Psi_n\\[1mm]
    \tilde\psi_n\end{pmatrix}
 =\begin{pmatrix}
  \dfrac{\la^2}{\mu^2\SP<F_n,F_n>} & -\dfrac{2\la}{\mu}(F^{-1}_n)^\top & 1 \\
   \dfrac{\la}{\mu}F^{-1}_n
     & \Bigl(\dfrac{\la^2}{\mu^2}-1\Bigr)I
      -\dfrac{2\la^2}{\mu^2}F^{-1}_nF^\top_n & \dfrac{\la}{\mu}F_n \\
   1 & -\dfrac{2\la}{\mu}F^\top_n & \dfrac{\la^2}{\mu^2}\SP<F_n,F_n>
 \end{pmatrix}
 \begin{pmatrix}\psi_{n-1}\\[1mm] \Psi_n\\[1mm] \psi_n\end{pmatrix}
\end{equation}
maps the general solution of these systems into the solution of the systems
of the same form, with the original and transformed potentials related by
equations
\begin{equation}\label{V1.VVFF}
 \mu V_n^{-1}=F_n+F^{-1}_{n+1},\qquad \mu\tilde V^{-1}_n=F_{n+1}+F^{-1}_n.
\end{equation}
\end{statement}

The expanded form of relations (\ref{V1.VVFF}) is
\begin{equation}\label{V1.VVFF'}
\begin{aligned}
 V_n&=\mu\frac{F_{n+1}+\SP<F_{n+1},F_{n+1}>F_n}
       {1+2\SP<F_n,F_{n+1}>+\SP<F_n,F_n>\SP<F_{n+1},F_{n+1}>},\\
 \tilde V_n&=\mu\frac{F_n+\SP<F_n,F_n>F_{n+1}}
       {1+2\SP<F_n,F_{n+1}>+\SP<F_n,F_n>\SP<F_{n+1},F_{n+1}>}.
\end{aligned}
\end{equation}
Each of these transformations is the substitution to the lattice
(\ref{V1.Vx}) from the lattice
\[
 F_{n,x}=\mu^2(F_{n-1}+F^{-1}_n)^{-1}-\mu^2(F_{n+1}+F^{-1}_n)^{-1}.
\]
It is easy to see that these formulae turn into (\ref{sc.vvff}) and
(\ref{sc.fx}) in the scalar case at $a=0$.

\begin{figure}[t]
\centerline{
\includegraphics[width=7cm]{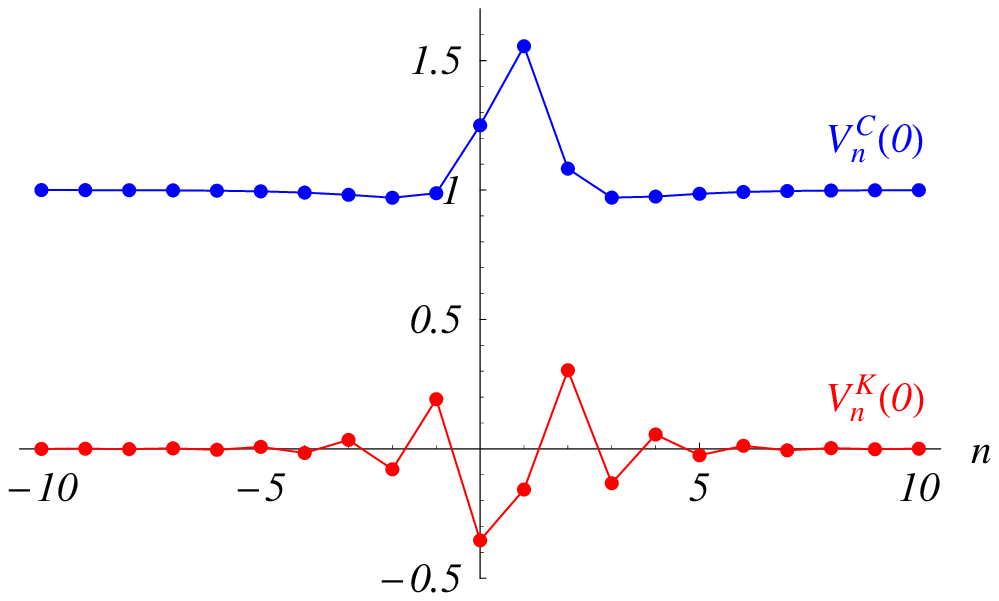}\qquad
\includegraphics[width=7cm]{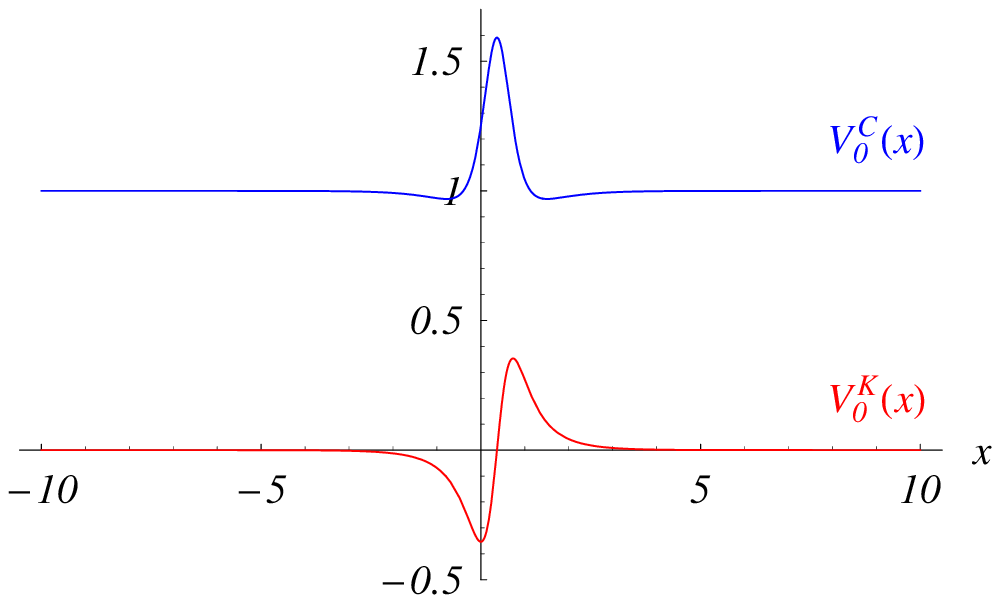}}
\centerline{
\includegraphics[width=7cm]{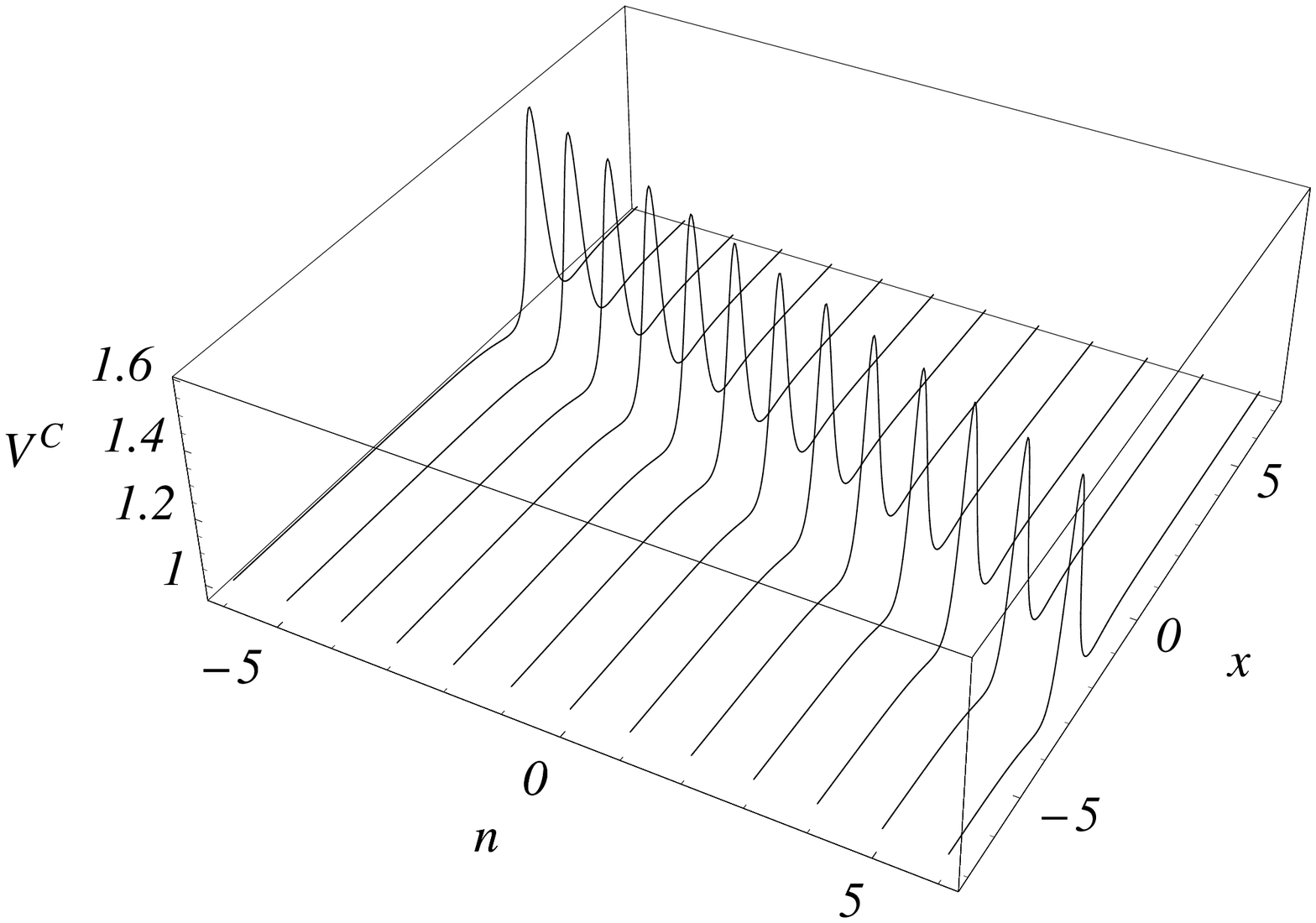}\qquad
\includegraphics[width=7cm]{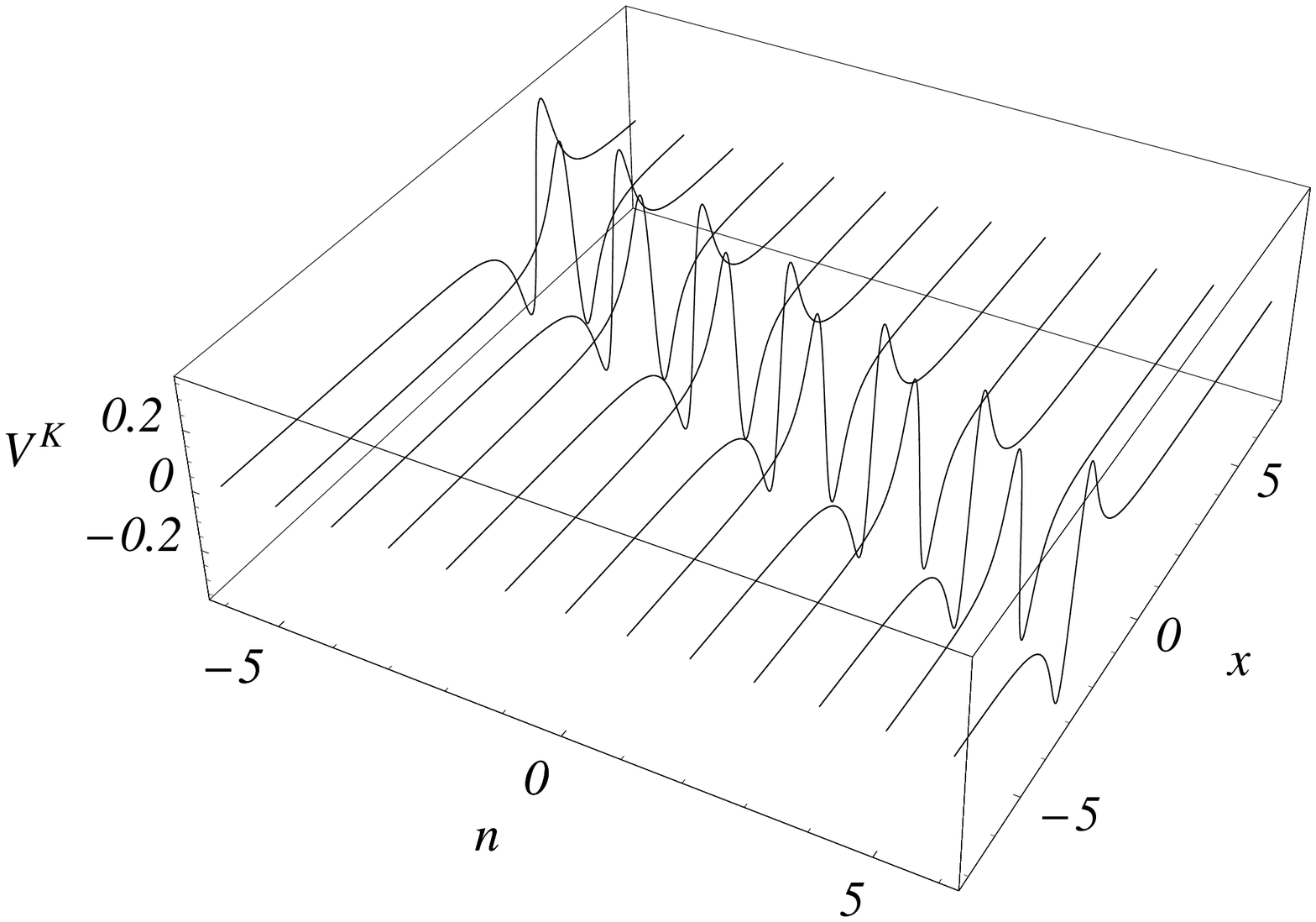}}
\caption{A soliton of the lattice (\ref{V1.Vx}); $C=(1,0)$, $K=(0,1)$, $\g=2$,
$c_1=1$, $c_3=-1$.}
\label{fig:V1.soliton}
\end{figure}

The derivation of the nonlinear superposition principle is not more
difficult than in the scalar case. One comes by multiplying the matrices $M$
of the form (\ref{V1.DT}) to the following Yang-Baxter mapping (cf eq.
(\ref{sc.R}) at $a=0$)
\begin{gather}
\nonumber
 F^{(j,k,\sigma)}_n=R(F^{(j,\sigma)}_n,F^{(k,\sigma)}_n;\mu^{(j)},\mu^{(k)}),\qquad
 F^{(k,j,\sigma)}_n=R(F^{(k,\sigma)}_n,F^{(j,\sigma)}_n;\mu^{(k)},\mu^{(j)}),\\[0.5em]
\label{V1.R}
 R(F,G;\mu,\nu)=
 \frac{(\mu^2-\nu^2)\SP<G,G>F+\mu(\nu\SP<F,F>-2\mu\SP<F,G>+\nu\SP<G,G>)G}
 {\SP<G,G>\SP<\nu F-\mu G,\nu F-\mu G>}.
\end{gather}

It is possible to obtain the analog of equation (\ref{sc.0NSP}), too. Let us
introduce the new vector variable $Z_n$ accordingly to the formulae
\[
 F_n=\mu(\tilde Z_n-Z_{n-1}),\qquad V^{-1}_n=Z_{n+1}-Z_{n-1}.
\]
Equations (\ref{V1.VVFF}) become equivalent to the single equation
\[
 \tilde Z_{n+1}-Z_n=\mu^{-2}(Z_{n+1}-\tilde Z_n)^{-1}
\]
under this change. Next, consider Darboux transformation corresponding to
the spectral value $\la=\nu$:
\[
 \hat Z_{n+1}-Z_n=\nu^{-2}(Z_{n+1}-\hat Z_n)^{-1}.
\]
The direct calculation shows that the repeated Darboux transformations
coincide: $\hat{\tilde Z}_n=\tilde{\hat Z}_n$ and moreover, the result is
given by the equation
\[
 \hat{\tilde Z}_n-Z_n=(\mu^{-2}-\nu^{-2})(\hat Z_n-\tilde Z_n)^{-1}.
\]
Iterations of the Darboux transformation are governed by the 3D-consistent
discrete equation on the square grid (the subscript $n$ is dummy):
\[
 Z^{(j+1,k+1)}_n-Z^{(j,k)}_n=\bigl((\mu^{(j)})^{-2}-(\nu^{(k)}\bigr)^{-2})
  \bigl(Z^{(j,k+1)}_n-Z^{(j+1,k)}_n\bigr)^{-1}.
\]
This equation with important applications in the discrete geometry was
introduced in \cite{Schief} (a special reduction was considered in
\cite{Adler_1995}), see also \cite{Bobenko_Suris_2002,Bobenko_Suris_2005}.

\begin{figure}[t]
\centerline{
\includegraphics[width=7cm]{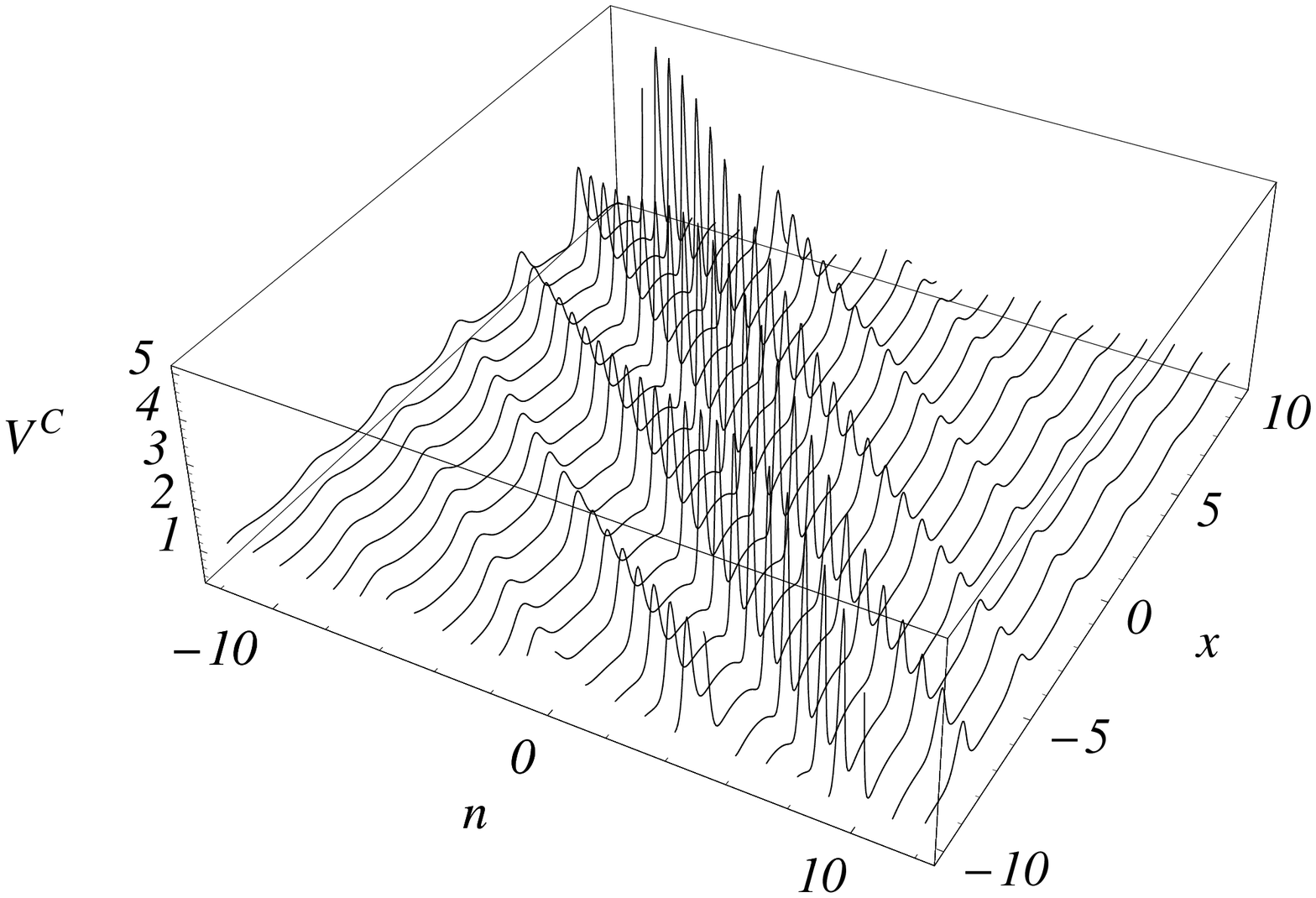}\qquad
\includegraphics[width=7cm]{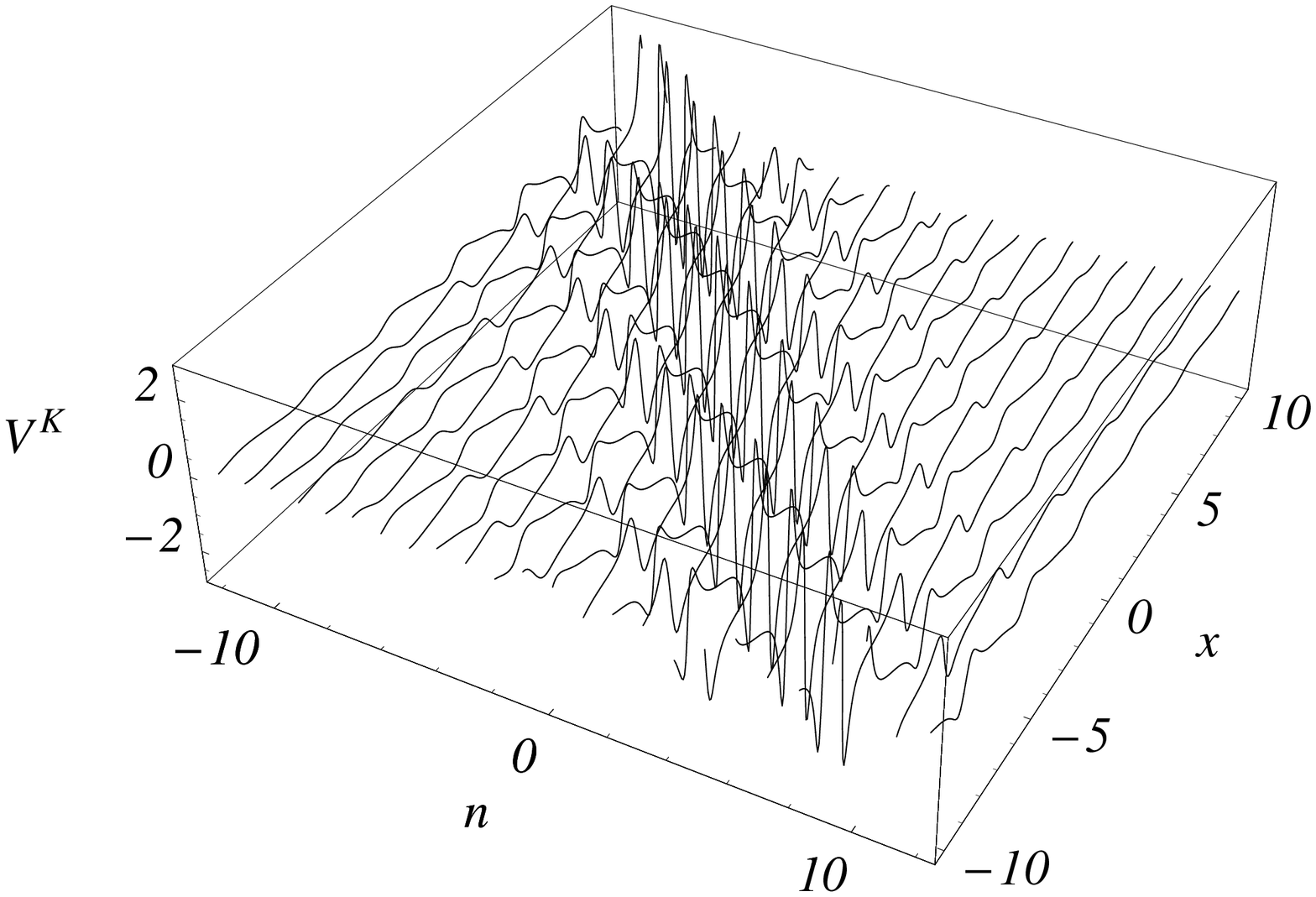}}
\caption{A breather of the lattice (\ref{V1.Vx}); $C=(1,0)$, $K=(0,1)$,
 $\g^{(1)}=\bar\g^{(2)}=0.5+i$, $c^{(1)}_1=c^{(2)}_1=0.5$,
 $c^{(1)}_3=c^{(2)}_3=1$.}
\label{fig:V1.breather}
\end{figure}

Let us make use of Darboux transformation for construction of the soliton
solution. The solution of the linear equations (\ref{V1.T}), (\ref{V1.D})
with constant coefficients $V_n=C=\const$, $\SP<C,C>=1$, at $\la=\mu$ reads
\begin{equation}\label{phi}
\begin{gathered}
 \phi_n=c_1\g^ne^{(\g-\g^{-1})x}+c_2\g^{-n}e^{(\g^{-1}-\g)x}+2c_3,\\
 \Phi_n=(-1)^nK+\mu\left(\frac{c_1\g^n}{1+\g}\,e^{(\g-\g^{-1})x}
    +\frac{c_2\g^{-n}}{1+\g^{-1}}\,e^{(\g^{-1}-\g)x}+c_3\right)C,\\
 \mu^2=\g+2+\g^{-1},\qquad
 \SP<C,K>=0,\qquad
 \g\SP<K,K>=(1-\g)^2(c_1c_2-c^2_3)
\end{gathered}
\end{equation}
(the latter relation is equivalent to the constraint $J=0$; and we do not
consider the cases of multiple eigenvalues $\mu=0$, $\mu=\pm2$). The
equations (\ref{V1.VVFF}) bring, after elementary transformations, to the
one-soliton solution of the lattice (fig.~\ref{fig:V1.soliton})
\[
 \tilde V_n=\frac{\phi_{n+1}\Phi_n+\phi_{n-1}\Phi_{n+1}}{\mu\phi^2_n}.
\]
Clearly, this solution always lies in the plane of the vectors $C,K$, that
is it is actually 2-component, independently on the dimension of the vector
space under consideration. The $C$-component is a soliton on the unit
background. Its shape is slightly different for positive and negative values
of $c_3$. The $K$-component has localized oscillations on the zero
background. They originate from the powers of $-1$ in the solution
(\ref{phi}) rather than a pair of complex conjugated points of the discrete
spectrum, that is this solution is not a genuine breather. However, the
additional dimension makes the breathers possible as well, in spite of the
absence of the parameter $a$ (fig.~\ref{fig:V1.breather}). The $N$-soliton
solution is constructed starting from a set of solutions of the form
(\ref{phi}) and evolves in the space spanned over the vectors
$C,K^{(1)},\dots,K^{(N)}$.

\section{Second vector generalization}\label{s.V2}

The lattice
\begin{equation}\label{V2.Ux}
 U_{n,x}=(\SP<U_n,U_n>+a)(U_{n+1}-U_{n-1})
\end{equation}
looks more natural and simple generalization of the lattice (\ref{sc.vx}).
In contrast with (\ref{V1.Vx}) it is integrable at arbitrary value of
parameter $a$, though it turns out to be not so important as in the scalar
case. Indeed, it can be easily eliminated or, more precisely, ``confined
inside the lattice'' at the expense of increasing by 1 the dimension of the
vector space under consideration. This is done by means of the orthogonal
complement: let $U_n$ be a solution of the lattice (\ref{V2.Ux}) then the
vector
\begin{equation}\label{V2.E}
 V_n=U_n+E,\quad E=\const,\quad \SP<U_n,E>=0,\quad \SP<E,E>=a
\end{equation}
(if $a<0$ then a pseudoeuclidean scalar product is used) satisfies the
lattice
\begin{equation}\label{V2.Vx}
 V_{n,x}=\SP<V_n,V_n>(V_{n+1}-V_{n-1}).
\end{equation}
This transformation does not lead to any problem when constructing solutions
since the reduction (\ref{V2.E}) is consistent with higher symmetries and
B\"acklund transformation. On the other hand, all formulae simplify
essentially (cf e.g. equations (\ref{V2.VVFF}) and (\ref{V2.UUFF}) below).
The matrices of the zero curvature representation become simpler as well.

The lattice (\ref{V2.Vx}) is the compatibility condition of the linear
systems
\begin{equation}
\label{V2.T}
 T\begin{pmatrix} \psi_{n-1} \\ \Psi_n \\ \psi_n \end{pmatrix}
  =\begin{pmatrix}
  0 & 0 & 1 \\[1em]
  0 & I -2V^{-1}_nV^\top_n & \la V^{-1}_n \\[1em]
  1 & -2\la(V^{-1}_n)^\top & \la^2/\SP<V_n,V_n>
 \end{pmatrix}
 \begin{pmatrix} \psi_{n-1} \\ \Psi_n \\ \psi_n \end{pmatrix},
\end{equation}
\begin{equation}
\label{V2.D}
 D_x\begin{pmatrix} \psi_{n-1} \\ \Psi_n \\ \psi_n \end{pmatrix}
  =\begin{pmatrix}
  -\la^2 & 2\la V^\top_{n-1} & 0 \\[1em]
  -\la V_n & 2V_nV^\top_{n-1}-2V_{n-1}V^\top_n & \la V_{n-1} \\[1em]
   0 & -2\la V^\top_n & \la^2
 \end{pmatrix}
  \begin{pmatrix} \psi_{n-1} \\ \Psi_n \\ \psi_n \end{pmatrix}.
\end{equation}
These systems possess the first integral (\ref{J}) in common, like
in the previous case.

\begin{statement}
Let $F_n=\Phi_n/\phi_n$ where $\psi=\phi$, $\Psi=\Phi$ is a particular
solution of the linear systems (\ref{V2.T}), (\ref{V2.D}) at $\la=\mu$ and
such that $J=\SP<\Phi_n,\Phi_n>-\phi_n\phi_{n-1}=0$. Then the transform
\begin{equation}\label{V2.DT}
 \begin{pmatrix}\tilde\psi_{n-1}\\[1mm] \tilde\Psi_n\\[1mm]
    \tilde\psi_n\end{pmatrix}
 =\begin{pmatrix}
  \dfrac{\la^2}{\mu^2\SP<F_n,F_n>} & -\dfrac{2\la}{\mu}(F^{-1}_n)^\top & 1 \\
  \dfrac{\la}{\mu}F^{-1}_n
   & \Bigl(1-\dfrac{\la^2}{\mu^2}\Bigr)I-2F^{-1}_nF^\top_n & \dfrac{\la}{\mu}F_n\\
   1 & -\dfrac{2\la}{\mu}F^\top_n & \dfrac{\la^2}{\mu^2}\SP<F_n,F_n>
 \end{pmatrix}
 \begin{pmatrix}\psi_{n-1}\\[1mm] \Psi_n\\[1mm] \psi_n\end{pmatrix}
\end{equation}
maps the general solution of these systems into solution of the systems of
the same kind, with the original and transformed potential related by the
equations
\begin{equation}\label{V2.VVFF}
 V_n=\mu\frac{F_{n+1}-\SP<F_{n+1},F_{n+1}>F_n}
   {1-\SP<F_n,F_n>\SP<F_{n+1},F_{n+1}>},\qquad
 \tilde V_n=\mu\frac{F_n-\SP<F_n,F_n>F_{n+1}}
   {1-\SP<F_n,F_n>\SP<F_{n+1},F_{n+1}>}.
\end{equation}
\end{statement}

In comparison with the previous Section, the equations (\ref{V2.VVFF}) are
slightly shorter than (\ref{V1.VVFF'}), but the analog of the lattice
(\ref{sc.fx}) is more cumbersome:
\[
 F_{n,x}=\frac{\mu^2\SP<F_n,F_n>
  \bigl((F_n-F^{-1}_{n-1})^{-1}-(F_n-F^{-1}_{n+1})^{-1}\bigr)}
  {\SP<(F_n-F^{-1}_{n-1})^{-1},F_n+F^{-1}_{n-1}>
   \SP<(F_n-F^{-1}_{n+1})^{-1},F_n+F^{-1}_{n+1}>}.
\]
The superposition of Darboux transformations is defined by the Yang-Baxter
map
\begin{gather}
\nonumber
 F^{(j,k,\sigma)}_n=R(F^{(j,\sigma)}_n,F^{(k,\sigma)}_n;\mu^{(j)},\mu^{(k)}),\qquad
 F^{(k,j,\sigma)}_n=R(F^{(k,\sigma)}_n,F^{(j,\sigma)}_n;\mu^{(k)},\mu^{(j)}),\\[0.5em]
\label{V2.R}
 R(F,G;\mu,\nu)=
 \frac{(\nu^2-\mu^2)\SP<G,G>F+\nu(\mu\SP<F,F>-2\nu\SP<F,G>+\mu\SP<G,G>)G}
 {\SP<G,G>\SP<\nu F-\mu G,\nu F-\mu G>}.
\end{gather}
Notice that the formulae (\ref{V1.R}) and (\ref{V2.R}) coincide up to the
permutation of $\mu$ and $\nu$ in the numerator. Despite of such similarity,
an analog of equation (\ref{sc.0NSP}) is probably lacked in this case.

\begin{statement}\label{th.noa}
The Darboux transformation is consistent with the reduction (\ref{V2.E}).
\end{statement}
\begin{proof}
Let us apply the change $V_n=U_n+E$, $F_n=H_n+h_nE$, $\SP<H_n,E>=0$ to the
equations (\ref{V2.VVFF}), with the scalar factor $h_n$ unknown for the
moment:
\begin{equation}\label{V2.h}
\begin{aligned}
 U_n+E&=\mu\frac{H_{n+1}+h_{n+1}E-(\SP<H_{n+1},H_{n+1}>+ah^2_{n+1})(H_n+h_nE)}
   {1-(\SP<H_n,H_n>+ah^2_n)(\SP<H_{n+1},H_{n+1}>+ah^2_{n+1})},\\
 \tilde U_n+E&=\mu\frac{H_n+h_nE-(\SP<H_n,H_n>+ah^2_n)(H_{n+1}+h_{n+1}E)}
   {1-(\SP<H_n,H_n>+ah^2_n)(\SP<H_{n+1},H_{n+1}>+ah^2_{n+1})}.
\end{aligned}
\end{equation}
Collecting the coefficients of $E$ yields the coupled algebraic equations for
$h_n$ and $h_{n+1}$. It is not obvious beforehand that their solution is
compatible with the shift in $n$. If this would be not the case then the
change $F_n=H_n+h_nE$ would be incorrect. However, the direct computation
proves that $h_n$ is defined by one and the same formula for all $n$ as a
solution of quadratic equation
\begin{equation}\label{V2.hH}
 ah^2_n-\mu h_n+\SP<H_n,H_n>+1=0,
\end{equation}
therefore the $E$-component is detached in the transformation
(\ref{V2.VVFF}).
\end{proof}

Statement \ref{th.noa} makes abundant the separate study of the case
$a\ne0$. Nevertheless, all formulae can be, in principle, rewritten for this
case as well, moreover, their rational structure can be preserved by use of
the stereographic projection for the quadric (\ref{V2.hH}):
\[
 H_n=\frac{(\nu^2-a)F_n}{\nu^2+a\SP<F_n,F_n>},\quad
 h_n=\frac{\nu(1+\SP<F_n,F_n>)}{\nu^2+a\SP<F_n,F_n>},\quad
 \mu=\nu+\frac{a}{\nu}
\]
(more rigorously, some other letter should be used here instead of $F$, but
we hope it will not lead to misunderstanding). For instance, the
substitution into (\ref{V2.h}) brings, under this parametrization, to the
B\"acklund transformation for the lattice with parameter (\ref{V2.Ux}):
\begin{equation}\label{V2.UUFF}
\begin{aligned}
 U_n&=\frac{(\nu^2+a\SP<F_n,F_n>)F_{n+1}-(a+\nu^2\SP<F_{n+1},F_{n+1}>)F_n}
   {\nu(1-\SP<F_n,F_n>\SP<F_{n+1},F_{n+1}>)},\\
 \tilde U_n&=\frac{(\nu^2+a\SP<F_{n+1},F_{n+1}>)F_n-(a+\nu^2\SP<F_n,F_n>)F_{n+1}}
   {\nu(1-\SP<F_n,F_n>\SP<F_{n+1},F_{n+1}>)}.
\end{aligned}
\end{equation}
The Yang-Baxter map (\ref{V2.R}) can be rewritten in more general form in a
similar way. The transformation (\ref{V2.UUFF}) turns into (\ref{V2.VVFF})
at $a=0$, while in the scalar case we come back to the transformation
(\ref{sc.vvff}), under identifying $U,F,\nu$ with $v,f,\mu$ respectively.

\begin{figure}[t]
\centerline{
\includegraphics[width=7cm]{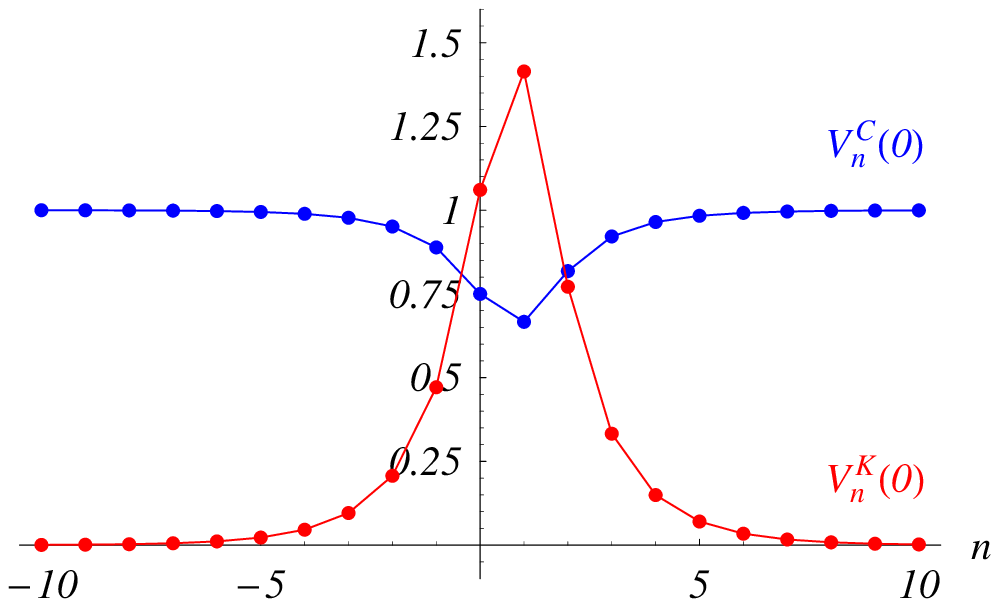}\qquad
\includegraphics[width=7cm]{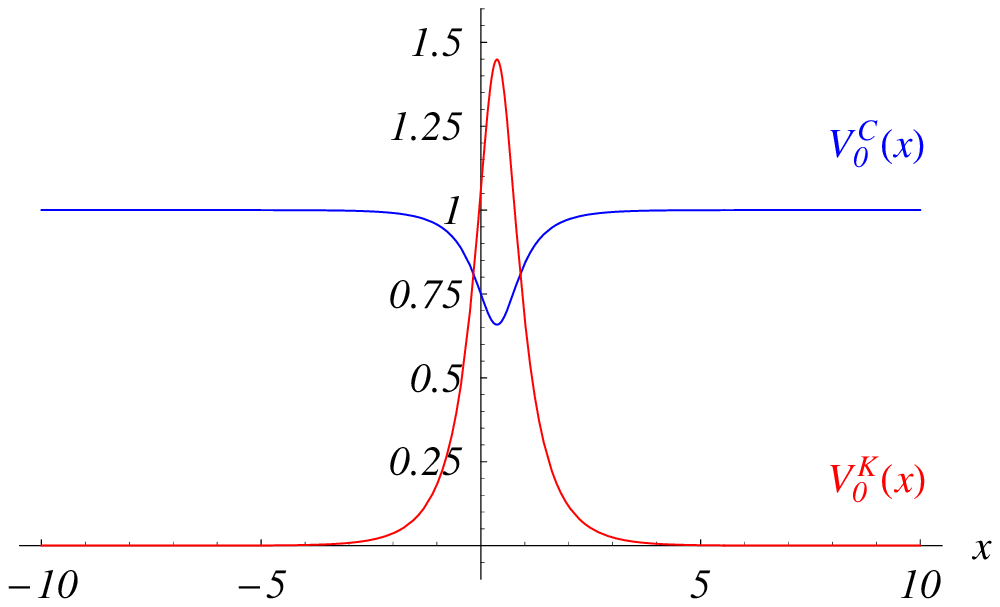}}
\centerline{
\includegraphics[width=7cm]{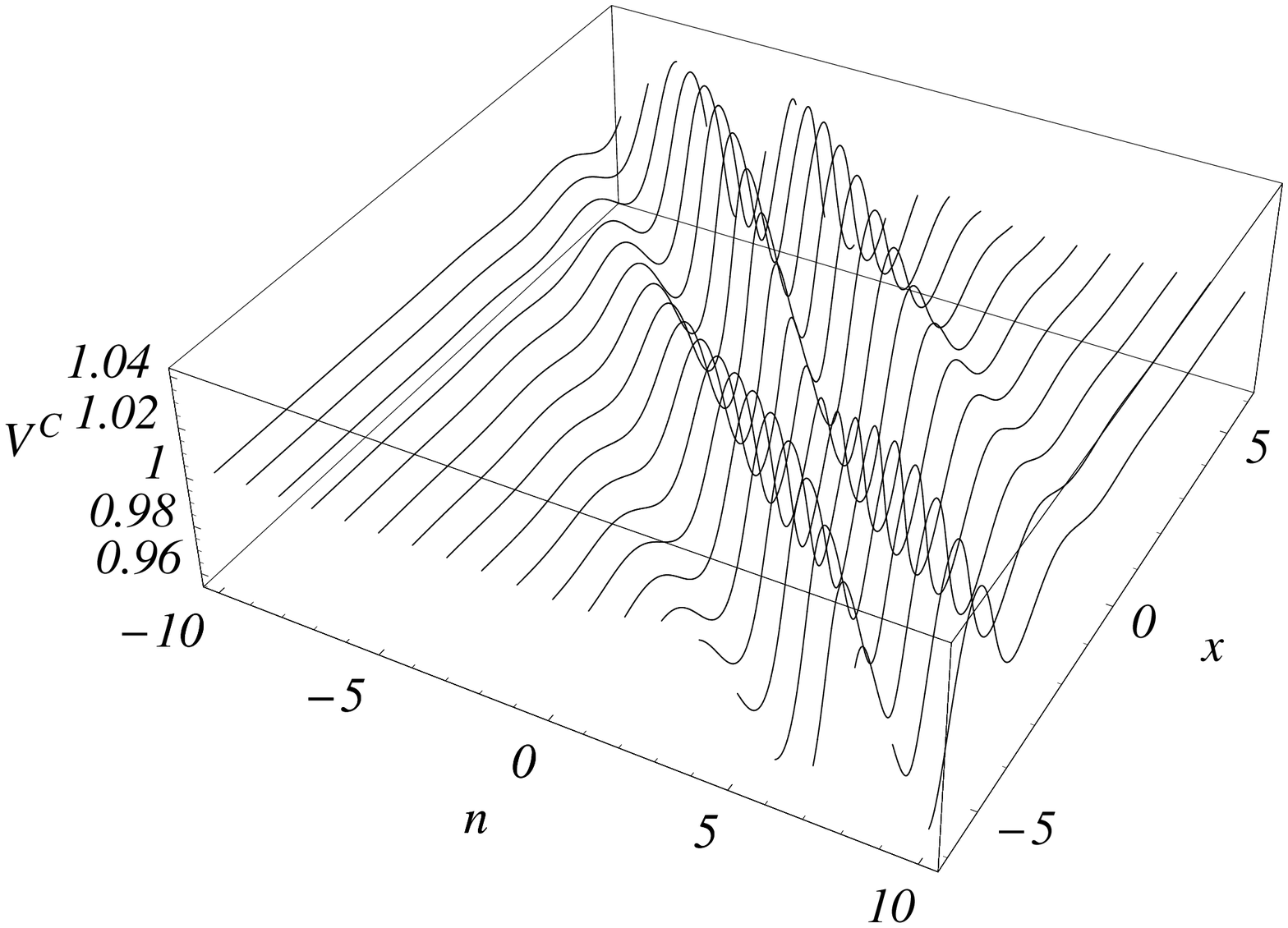}\qquad
\includegraphics[width=7cm]{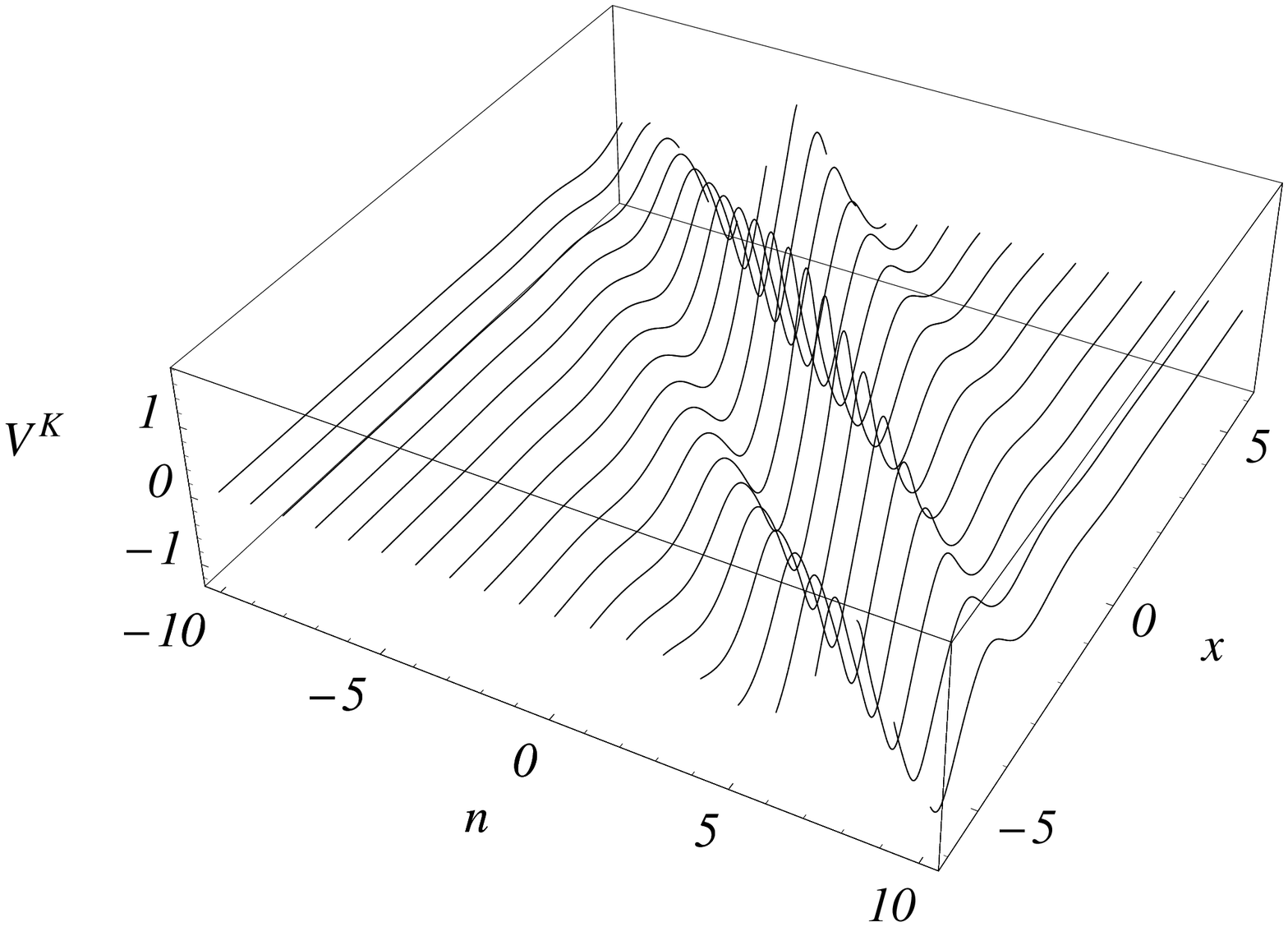}}
\caption{Solutions of the lattice (\ref{V2.Vx}) at $C=(1,0)$, $K=(0,1)$.
 Soliton: $\g=2$, $c_1=1$, $c_3=-1$;
 breather: $\g^{(1)}=\bar\g^{(2)}=1+1.5i$, $c^{(1)}_1=c^{(2)}_1=3$,
 $c^{(1)}_3=c^{(2)}_3=-1$.}
\label{fig:V2}
\end{figure}

It is not difficult to compute, by use of (\ref{V2.VVFF}), the one-soliton
solution
\[
 \tilde V_n=\mu\frac{\phi_{n+1}\Phi_n-\phi_{n-1}\Phi_{n+1}}
  {\phi_n(\phi_{n+1}-\phi_{n-1})},
\]
where the solution $\phi_n,\Phi_n$ of equations (\ref{V2.T}), (\ref{V2.D})
with constant coefficients is given by almost the same formulae (\ref{phi})
as before, with the only difference that the factor $(-1)^n$ in front of $K$
disappears. This distinction leads to the absence of blinking oscillations
in the one-soliton solution which is more natural from the point of view of
the continuous limit. In all other respects the construction of multisoliton
and breather solutions is analogous to the previous case.

\section{Higher symmetries and associated systems}\label{s.sym}

Both vector lattices (\ref{V1.Vx}) and (\ref{V2.Vx}) belong to an infinite
hierarchy of commuting flows. We restrict ourselves by consideration of the
simplest higher symmetries which are of the second order with respect to the
shift in $n$. In the scalar case one has, setting $a=0$ for simplicity, the
pair of consistent lattices
\begin{equation}\label{sc.vt}
 v_{n,x}=v^2_n(v_{n+1}-v_{n-1}),\quad
 v_{n,t}=v^2_n(v^2_{n+1}(v_{n+2}+v_n)-v^2_{n-1}(v_n+v_{n-2})).
\end{equation}
Obviously, the first of these equations can be solved with respect to
$v_{n+1}$ or $v_{n-1}$, and this allows to express recursively all $v_j$
through the pair of variables $u=v_{n+1}$, $v=v_n$. After this, the symmetry
takes the form of Kaup-Newell evolution system
\[
 u_t=u_{xx}+(2u^2v)_x,\quad v_t=-v_{xx}+(2uv^2)_x
\]
and the shift in $n$ defines an explicit auto-substitution for this system
(the simplest type of B\"acklund transforms).

The lattice (\ref{V1.Vx}) and its symmetry can be compactly written in the
form preserving the structure (\ref{sc.vt})
\begin{equation}\label{V1.t}
 V_{n,x}=P_{V_n}(V_{n+1}-V_{n-1}),\quad
 V_{n,t}=P_{V_n}(P_{V_{n+1}}(V_{n+2}+V_n)-P_{V_{n-1}}(V_n+V_{n-2}))
\end{equation}
by use of the operator $P_V(U)=2\SP<V,U>V-\SP<V,V>U$. It is easy to check
that the identity $(P_V)^{-1}=P_{V^{-1}}$ is valid for this operator. Making
use of it one can solve, like before, the first equation with respect to
$V_{n+1}$ or $V_{n-1}$ and to express all $V_j$ through the pair of
variables $U=V_{n+1}$, $V=V_n$. This brings the symmetry to the form of the
vector generalization of Kaup-Newell system
\begin{equation}\label{V1.UV}
 U_t= U_{xx}+(4\SP<U,V>U-2\SP<U,U>V)_x,\quad
 V_t=-V_{xx}+(4\SP<U,V>V-2\SP<V,V>U)_x.
\end{equation}
Analogously, the commuting flows for the second vector lattice are
\begin{equation}\label{V2.t}
\begin{aligned}
 V_{n,x}&= \SP<V_n,V_n>(V_{n+1}-V_{n-1}),\\
 V_{n,t}&= \SP<V_n,V_n>\bigl(\SP<V_{n+1},V_{n+1}>(V_{n+2}-V_n)
     +\SP<V_{n-1},V_{n-1}>(V_n-V_{n-2})\\
  &\qquad+2(\SP<V_{n+1},V_n>+\SP<V_n,V_{n-1}>)(V_{n+1}-V_{n-1})\bigr)
\end{aligned}
\end{equation}
and the associated evolution system reads
\begin{equation}\label{V2.UV}
 U_t= U_{xx}+4\SP<U,V>U_x+2\SP<U,U>V_x,\quad
 V_t=-V_{xx}+4\SP<U,V>V_x+2\SP<V,V>U_x
\end{equation}
which is another vector analog of Kaup-Newell system.

Another interesting type of associated systems is obtained for the scalar
quantities
\[
 p_n=\SP<V_n,V_n>,\quad q_n=2\SP<V_n,V_{n-1}>
\]
which satisfy, in virtue of any of the pair (\ref{V1.t}) or (\ref{V2.t}) one
and the same two-dimensional modified Volterra lattice
\begin{equation}\label{2DV}
 p_{n,t}+2p^2_n(p_{n+1}-p_{n-1})=p_n((q_{n+1}+q_n)_x+q^2_{n+1}-q^2_n),\quad
 p_{n,x}=p_n(q_{n+1}-q_n).
\end{equation}
It can be written in the form
\[
 p_{n,t}+2p^2_n(p_{n+1}-p_{n-1})=(r_np_n)_x,\quad
 (p_{n+1}p_n)_x=p_{n+1}p_n(r_{n+1}-r_n),
\]
as well, where $r_n=q_{n+1}+q_n=2\SP<V_n,V_{n+1}+V_{n-1}>$. These lattices
are closely related to Mikhailov lattices introduced in \cite{Mikhailov}.

The lattices (\ref{V1.t}), (\ref{V2.t}) can be effectively used for the
construction of particular solutions of the systems (\ref{V1.UV}),
(\ref{V2.UV}) and the lattice (\ref{2DV}). Along with the construction
method of the soliton-type solutions described above, one can use to this
end the periodic closure $V_{n+N}=CV_n$ with orthogonal operator $C$ which
leads to the finite-dimensional dynamical systems.

\section{Further vector analogs}\label{s.more}

Remind that the classification problem of scalar integrable lattices of
Volterra type was solved by Yamilov \cite{Yamilov_1983} within the symmetry
approach. Recently one of the authors has obtained an analogous
classification of the vector Volterra lattices on the sphere, that is under
the constraint $\SP<V_n,V_n>=1$ \cite{Adler_2008}. This constraint
essentially simplifies the problem which is very complicated and remains
open for the case of free space. Other simplifying assumptions can be used
of course, for example the polynomiality of the lattice. It should be noted
that in the continuous case very many polynomial equations are known; we
mention only the papers \cite{Fordy, Tsuchida_Wadati, Sokolov_Wolf} which
contain the examples and some classification results for the vector systems
of derivative nonlinear Schr\"odinger type, equations (\ref{V1.UV}),
(\ref{V2.UV}) being just two instances of such systems. In the discrete
case, however, the polynomiality is not too natural assumption, as one can
see already from the Yamilov list of scalar lattices. In the vector setting
we have not succeeded in finding another polynomial Volterra type lattices
possessing higher symmetries aside from (\ref{V1.Vx}), (\ref{V2.Ux}).

Our search of integrable lattices was based on the straightforward method of
undetermined coefficients. In the simplest case the lattice and its symmetry
are of the form
\[
 V_{n,x}=a^{(1)}V_{n+1}+a^{(0)}V_n+a^{(-1)}V_{n-1},\quad
 V_{n,t}=b^{(2)}V_{n+2}+\dots+b^{(-2)}V_{n-2}
\]
where the scalar coefficients $a^{(i)}$ are linear with respect to the
scalar products of $V_{n+1}$, $V_n$, $V_{n-1}$ and $b^{(i)}$ are quadratic with
respect to the scalar products of $V_{n+2},\dots$, $V_{n-2}$. It is easy to
find that the homogeneous lattice contains 18 parameters and its symmetry
contains 600 ones. Calculating of the cross derivatives yields a system of
bilinear equations for the coefficients. Although this system is very bulky,
its solving is, in principle, not difficult since the equations are very
overdetermined and sparse (in particular, a large part of equations is
monomial). The answer is the consistent pairs of the lattices (\ref{V1.t})
and (\ref{V2.t}) (there are also few solutions with $a^{(1)}=a^{(-1)}=0$,
but all such lattices can be reduced to the scalar ones and therefore they
are not of interest for us).

It is clear that the scope of this method in this problem is very
restricted. If one takes the coefficients $a^{(i)}$ quadratic with respect
to the scalar products and $b^{(i)}$ of the fourth degree then the number of
unknown parameters in the lattice and its symmetry becomes 63 and 15300
respectively, and even the calculation of the commutator becomes not so
trivial task. This case is still manageable, but with the empty answer.

We also have partially analyzed the case when the lattice is of the second
order with respect to the shift in $n$ and its symmetry is of the fourth
order, that is
\[
 V_{n,x}=a^{(2)}V_{n+2}+\dots+a^{(-2)}V_{n-2},\quad
 V_{n,t}=b^{(4)}V_{n+4}+\dots+b^{(-4)}V_{n-4}.
\]
One may hope that some vector analogs of Narita-Bogoyavlensky lattice
\cite{Narita,Bogoyavlensky,Suris} appear here, more precisely, analogs of
some its modification with odd degree of nonlinearity, for example
\[
 v_{n,x}=v_n(v_{n+2}v_{n+1}-v_{n-1}v_{n-2}) \qquad \text{or}\qquad
 v_{n,x}=v_{n+1}v^3_nv_{n-1}(v_{n+2}v_{n+1}-v_{n-1}v_{n-2}).
\]
Notice that classification of such lattices is not known even in the scalar
case. Unfortunately, the analogs of Narita-Bogoyavlensky lattice have not
been discovered, however we have found two more lattices relative to
Volterra lattice:
\begin{gather}
\label{V3}
 V_{n,x}=\SP<V_n,V_n>\bigl(
   \SP<V_{n+1},V_{n+1}>(V_{n+2}+V_n)-\SP<V_{n-1},V_{n-1}>(V_n+V_{n-2})\bigr),\\
\label{V4}
 V_{n,x}=\SP<V_{n+1},V_n>\SP<V_n,V_{n-1}>(V_{n+2}-V_{n-2}).
\end{gather}
Each of these lattices possesses 4-th order symmetry which we do not bring
because of their length. The study of these examples falls beyond the scope
of our article. We only notice that the lattice (\ref{V3}) generalizes the
second flow of the modified Volterra lattice (\ref{sc.vt}), so that this flow
admits at least three vector analogs. The question on the number of vector
analogs for the higher flows of the hierarchy remains open. The lattice
(\ref{V4}) in the scalar case is a modification of the Volterra lattice on
the ``stretched'' grid:
\[
 v_{n,x}=v_{n+1}v^2_nv_{n-1}(v_{n-2}-v_{n-2})\quad
 \xrightarrow{~~u_n=v_{n+2}v_{n+1}v_nv_{n-1}~~}\quad
 u_{n,x}=u_n(u_{n-2}-u_{n-2}),
\]
but in the vector case this substitution makes no sense and the lattice
(\ref{V4}) seems to be an independent object. The zero curvature
representations and B\"acklund transformations for the lattices (\ref{V3}),
(\ref{V4}) are not known for now.

Summing up, we may say that the classification of the polynomial lattices of
Volterra and Narita-Bogoyavlensky types is a very difficult open problem,
probably with very scarce answers. The alternative approaches to the method
of undetermined coefficients are the analysis of the necessary integrability
conditions in the form of canonical conservation laws \cite{Yamilov_1983,
Levi_Yamilov, Yamilov_2006} and the perturbative approach
\cite{Mikhailov_Novikov}, however the contemporary state of the theory does
not allow to effectively apply them, even in the scalar case.

\paragraph{Acknowledgements.} We thank Ravil Yamilov and Yaroslav Pugai for many
fruitful discussions. The research of V.A. was supported by RFBR grants
06-01-92051-KE, 08-01-00453 and NSh-3472.2008.2.


\end{document}